# Single-laser pulse toggle switching in CoHo and CoDy single layer alloys : when domain wall motion matters

B. Kunyangyuen[1], G. Malinowski[1], D. Lacour[1], J.-X. Lin[1], Y. Le Guen[1], L. D. Buda-Prejbeanu[2], S. Mangin[1], J. Gorchon[1], J. Hohlfeld[1] and M. Hehn[1,*]

[1] Université de Lorraine, CNRS, IJL, F-54000 Nancy, France

[2] Univ Grenoble Alpes, CEA, CNRS, Grenoble INP, SPINTEC, 38000 Grenoble, France

* *michel.hehn @univ-lorraine.fr, gregory.malinowski@univ-lorraine.fr*

**Abstract**

Single pulse All Optical Helicity-Independent Toggle Switching is observed in CoHo and CoDy alloys single layers. An original reversal mechanism is reported which contrasts with those observed to date. It is shown that the reversal process is on the µs timescale involving the reorganization / coalescence of domains and domain walls. The toggle switching is explained considering the unbalanced magnetization distribution stabilized after the ultrashort laser pulse.

## Introduction

Ultrafast All Optical- Helicity Independent switching (AO-HIS) represents the ability to reverse the magnetization of a nanostructure without any applied magnetic field, only with the use of a single laser pulse. In some conditions, a characteristic time scale to cross zero magnetization of 1ps could be observed. This reversal process, the fastest ever reported for magnetic materials, appears to be of crucial importance for generating smaller, faster and frugal storage technologies. First observed in the $(FeCo)_xGd_{1-x}$ ferrimagnetic alloy [Sta06, Rad11, Ost12], AO-HIS has been shown to result from an ultrafast heating process combined with distinct magnetic moment dynamics of the rare earth and transitions metals elements and related to the high transient electron temperature being out of equilibrium with the lattice. The exchange-driven angular momentum transfer between the rare earth, Gd, and transition metals, Fe and Co, elements that are antiferromagnetically coupled results in the magnetization reversal of $(FeCo)_xGd_{1-x}$ alloy [Atx14, Dav20]. The same process has been evoked to explain the reversal in Gd/Co multilayers [Lal17] and in half-metallic ferrimagnetic Heusler alloys $Mn_2Ru_xGa_{1-x}$ which has two inequivalent Mn sublattices [Ban20]. The study of AO-HIS in other Co-RE alloys single layers reveals that only a partial single-pulse all-optical reversal could be achieved and only for the first few pulses in case of $Co_{0.75}Dy_{0.25}$ or $Co_{0.75}Tb_{0.25}$ films [Hu23]. Adding Gd in $Co_xDy_{1-x}$, $Co_xTb_{1-x}$ and $Co_xHo_{1-x}$ alloys led to AO-HIS while it fails when Gd concentration is reduced to zero [Zha24]. Finally, AO-HIS was observed in thin Co/Ho multilayers [Pen23]. The reversal process could not be identified but appears to be very slow. In this paper, we report on single pulse All Optical Helicity-Independent toggle switching in CoHo and CoDy alloys thin films and we shed light on the long timescale process leading to the magnetization reversal.

## Experimental details

The magnetic properties of very thin $Co_xHo_{1-x}$ films are not reported in literature even if, from 1975 to 1990, the magnetic and magneto-optic properties of Co-RE alloys have been widely studied for magneto-optic recording [Uch95, Han91, Han89]. Therefore, Glass/Ta(5)/Pt(5)/$Co_xHo_{1-x}$(*3*)/Ta(5)/Pt(1) multilayers (thickness in nm) were deposited by dc magnetron sputtering and their magnetic properties have been studied. Considering the work of P. Hansen et al. [Han91], Co atomic concentration *x* between 64 and 79% have been tested in order to cover the concentration for which magnetic compensation occurs at room temperature. The thickness of the $Co_xHo_{1-x}$ film was set to 3nm to fulfil the domain size criterion for the observation of all-optical switching in magnetic thin films [Had16] i. e. that the thin film hosts minimum size for domains to be able to stabilize AOS switched domains. The magneto-optical Kerr hysteresis loop have been measured at room temperature in the polar geometry. All the samples have perpendicular to film magnetization (PMA) and exhibit high squareness of the hysteresis loops with 100% remanence (Figure S1). All magneto-optical hysteresis cycles have the same polarity, which means that it is always the same sublattice that dominates the magnetization (has a larger moment). In other words, we do not see a magnetic compensation in the measured concentration range. Because of the positive sign of the Kerr rotation at high positive field, we attribute the dominant sublattice to the transition metal and so the alloy is always dominated by the Co sublattice. In the case of

CoxHo1-x films that are 30 nm thick, the magnetic compensation is found to be around 74.5% [Boo25]. As already observed in CoFeGd alloys [Bel22], we ascribed the shift of the magnetic compensation in 3 nm thick films to the polarization induced in Pt by the Co subnetwork.

### Optical switching of CoxHo1-x alloys

Optical switching and TR-MOKE dynamic experiments have been performed using a Ti:sapphire femtosecond-laser source and regenerative amplifier. In case of single pulse experiments, the resulting domain structure was observed several second after the laser pulse. In case of TR-MOKE measurements, a delay line up to 1.2 ns was used to delay the 400 nm probe pulse relatively to the 800 nm pump pulse. In this case, the repetition rate of the femtosecond laser is 5 kHz and a magnetic field is applied along the out-of-plane direction to reset the magnetization. A nice and reproducible reversal could be observed for films with Co concentration between 74.8 and 79% (Figure S2). Those concentrations are equal and slightly higher than the room temperature magnetization compensation of the $Co_xHo_{1-x}$ alloy in the thick film [Boo25]. In the case of 75.7%, a perfect single pulse switching is observed at fluence of 11 mJ/cm$^2$ and pulse duration of 50 fs, over 30000 pulses (Figure 1). This constitutes the first demonstration of full toggle reversal of CoxHo1-x alloys. This extends the property obtained in the Co/Ho multilayer to CoHo alloys and is counterintuitive as discussed in [Pen23]. Indeed, with the use of Ho instead of Gd, we reinforced the spin orbit coupling and the total magnetic momentum with respect to Tb and Dy, we expected to strengthen the characteristics of Tb- and Dy-based compounds for which only a partial single-pulse all-optical switching was observed [Hu23].

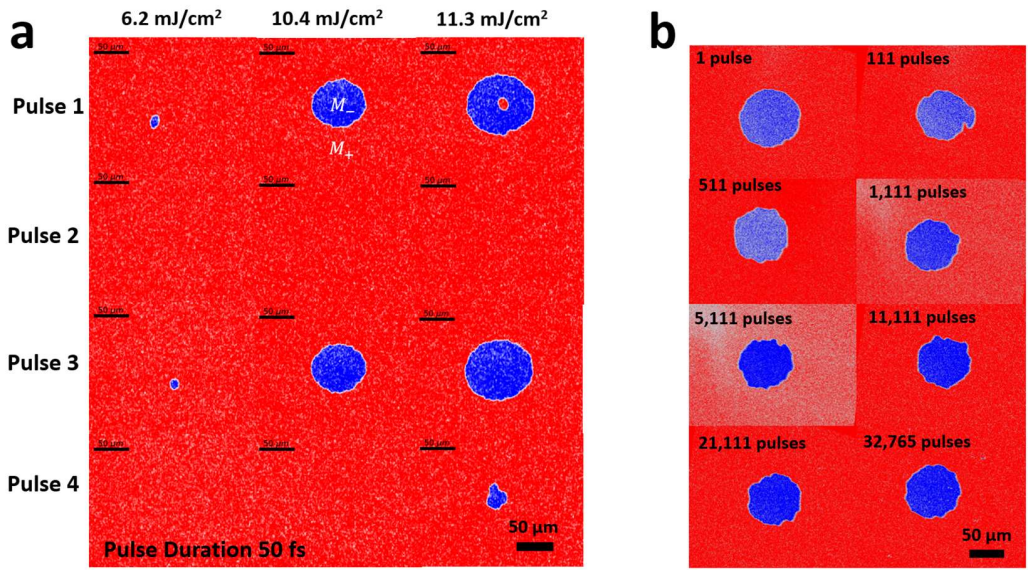

**Figure 1:** - Magneto-optical Kerr microscopy images obtained on a Glass/Ta(5)Pt(5)(Co75.7Ho24.3)(3)Ta(5)Pt(1) thin film. (a) Single pulse experiments for 4 successive pulses, increasing the fluence from the left to the right column; (b) Single pulse experiments done for a fluence of 9 mJ/cm$^2$ and a pulse duration of 50 fs and resulting domain structure after 1, 111, 511, 1111, 21111 and 32765 pulses. Each pulse induces a toggle switching.

One way to get more insights into the reversal process consists in checking (i) the pulse duration versus fluence state diagram and (ii) the dynamics of the reversal that are both reported in figure 2. The state diagram is at first glance, very similar to the one of FeCoGd alloys [Wei21]. The critical fluence to switch the magnetization, $F_{SW}$ (black curve in Fig. 2a), increases with pulse duration while the critical fluence to end in a multiple domain state, $F_{MD}$ (red curve in Fig. 2a) remains mostly constant. For a typical fluence of 10.5 mJ/cm$^2$, a mostly circular domain is written for pulse durations up to $\tau_{max}$ around 1000 fs beyond which the shape of the magnetic domains becomes increasingly ill-defined, even dendritic, with a continuous decrease of domain size (see figure 2(c)). At this transition, which is not well defined in pulse duration, the toggle switching is continuously lost (see figure S3). For fluences above $F_{MD}$ (around 11 mJ/cm$^2$), a multidomain state develops at the center of the laser beam while toggle switching is kept in the outer part of the laser beam where the fluence is the lowest thanks to the pump beam gaussian intensity profile. The two regions are delimited by dotted white lines in the two first domains images of figure 2(c). Here again, the transition is not well defined in pulse duration leading to dispersed experimental points. The reversal dynamics has been studied and is reported in Figure 2b. Surprisingly, the reversal of Co sublattice which is probed in our setup, could not be observed with our 1.2 ns delay line. On the contrary, an ultrafast demagnetization is measured with a typical demagnetization timescale below 1 ps and a recovery occurs. The recovery speed depends on the magnitude of the applied field. During the demagnetization and remagnetization, the magnetization does not precess around or relax to an in-plane direction. This means the PMA is always maintained and no re-orientation of the magnetic anisotropy is present.

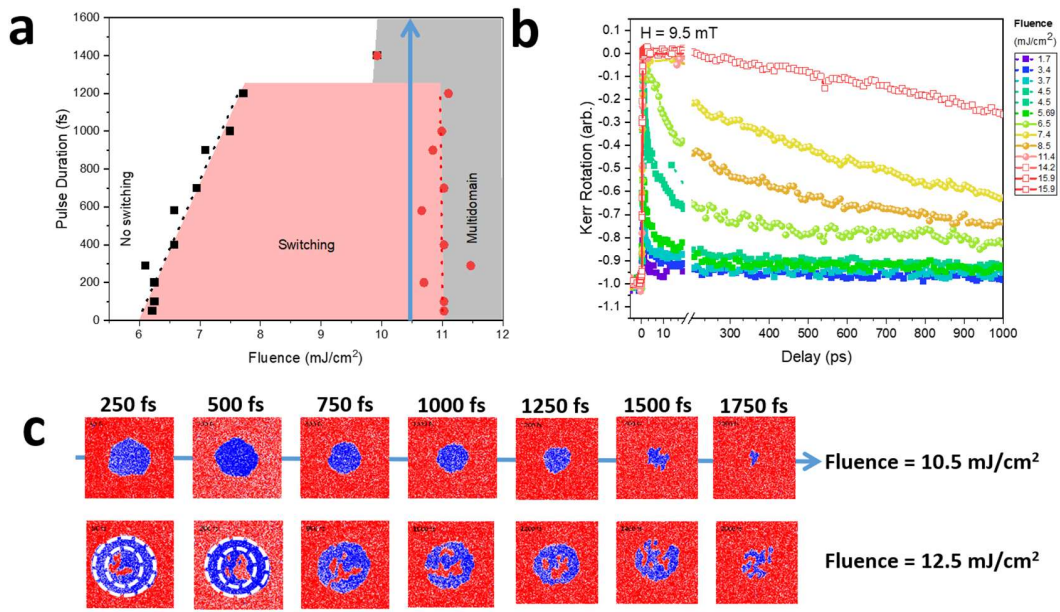

**Figure 2:** Measurements done on a Glass/Ta(5)Pt(5)(Co75.7Ho24.3)(3)Ta(5)Pt(1) thin film (a) Laser pulse duration versus fluence State diagram. Black curve : critical fluence to switch the magnetization, $F_{SW}$; Red curve, critical fluence to end in a multiple domain state, $F_{MD}$. (b) Magnetization dynamics versus time for an applied field of 9.5 mT and a pulse duration of 50 fs. (c) Evolution of the domain configuration for fluences of 10.5 mJ/cm$^2$ (blue arrow) and 12.5 mJ/cm$^2$. For fluence of 12.5 mJ/cm$^2$, dotted white lines are added in the two first domains images to delimit regions of toggle switching and multidomain at the center of the laser beam.

For fluences above 6.2 mJ/cm$^2$, a full demagnetized state is reached that recovers slowly with time. From those measurements, a clear correlation could be drawn: for fluence values such that total demagnetization is achieved, a reversal is obtained; if full demagnetization is not achieved, the reversal does not occur. On the other hand, if the fluence is too high, a multidomain state is stabilized. At this stage, our time-resolved setup prevents further analysis of the long time stabilized demagnetized state being impossible to know the features of the magnetic state (size of the domains, remagnetization process, etc...). An idea of the reversal time scale is needed.In order to overcome the limit imposed by the delay line and obtain a deeper understanding of the magnetization dynamics after 1 ns, Hall crosses have been fabricated from our thin films and long time electrical measurements have been performed. A special care has been devoted to the thermal effect during lithography and etching process to avoid changes in the magnetic properties of the alloys. Details on the process are given in the supplementary materials. As can be seen in figure 3a, a clear reversal of the magnetization can be observed on a time scale of 1 µs, and the reversal tail extends to around 300 µs. A second pulse causes the reversal back to the initial state with the same critical time reversal. Pulses 3 and 4 confirms the trend. The magnetic state after reversals have been imaged using Kerr microscopy and a full reversal is indeed observed (Fig. 3b). The origin of this slow reversal process can only be linked to a domain wall dynamical process. A same conclusion was drawn in the case of Co/Pt with close dynamical time scales [Vom17]. At the time, "further theoretical investigations were invoked including thermally induced nucleation/propagation process with the constraint of a static magnetization surrounding the laser spot (or thermal spot)".

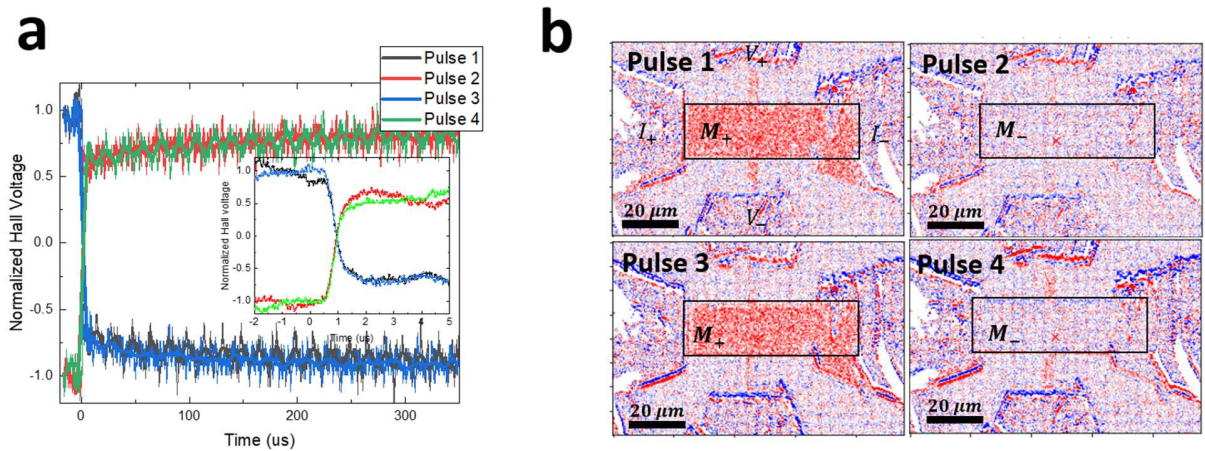

**Figure 3:** Experiments done on Si/SiO(500)/Ta(5)Pt(5)(Co76.5Ho23.5)(3)Ta(5)Pt(1). (a) Normalized Hall voltage versus time after pulse 1, 2, 3 and 4. Inset : zoom on the first µs time scale. Pulse conditions : fluence 5 mJ/cm2 and pulse duration 50fs. (b) Kerr microscope image measured after the laser pulses. Initial magnetic state is M- that has been subtracted with the background before the first pulse. It corresponds to the white contrast. Toggle switching lead to the appearance of the red M+ contrast. The fluence used to get the reversal is slightly less then in figure 2 because of the use of Si/SiO substrate instead of Glass.

Aside from applying magnetic fields, magnetic domains and domain walls have been shown to be deterministically manipulated using different external stimuli like electric fields [Sch12], lattice strain [Lei13], electric currents [Yam04,

Tor12], as well as thermal gradients [Tor12, Hin11] induced by localized femtosecond laser pulses [San16, Que18]. In order to identify which process is underway here, we carried out the experiment reported in figure 4. Starting from a sample with a saturated magnetization M+ (red color in figure 4), a first laser pulse leads to the appearance of a reversed domain with magnetization M- corresponding to the blue domain. The laser beam center is then moved to the black cross and a second pulse, that overlaps the previous written domain, is shined on the sample. We clearly see a toggle reversal: the part of domain that was reversed switched back while the part of the back ground which was not reversed changed its state towards blue. A third pulse at the same position brings the magnetic state to the one after the first pulse. It appears clearly that neither the domain configuration through its stray field nor the temperature gradient induced by the laser pulse are playing the key role. A very local process has to be invoked.

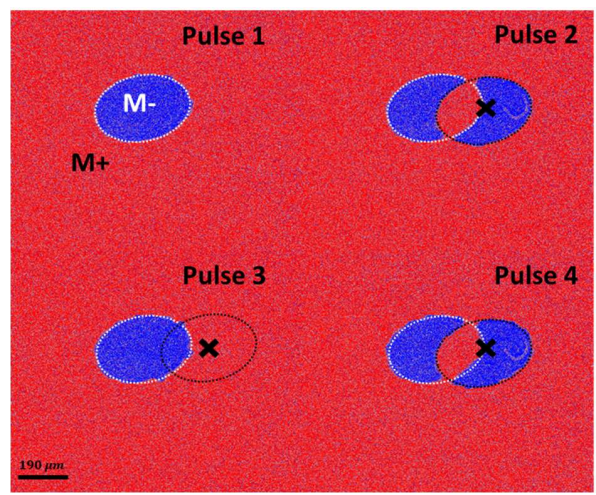

**Figure 4:** Sequential pulses to exclude thermal gradient induced or stray field induced reversal. The origin of the process is local. Experiments done on Glass/Ta(5)Pt(5)(Co75.7Ho24.3)(3)Ta(5)Pt(1) with pulse duration of 50 fs and fluence of 9 mJ/cm$^2$. Details in the text.

It is now well known that demagnetization of a ferro or ferrimagnetic layer induces the creation of spin current at very short time scales or lead to a transfer of angular momentum between the two sub lattices of a ferrimagnet [Rad11, Ost12]. We also know that transition metals and rare earth have different demagnetization time. Using element sensitivity of x-ray magnetic circular dichroism, it was indeed shown that typical demagnetization time of Co in the alloy was around 200 fs [Lop13, Ber14] while the demagnetization of the rare-earth is systematically slower : 480 fs [Lop13, Ber14] in case of Gd, 280-500 fs [Lop13, Ber14] in case of Tb, 610-980 fs [Abr21, Fer17] in case of Dy. No data are unfortunately available in case of Ho to the best of our knowledge. We will make the hypothesis that demagnetization time is longer in Ho that in Co, with a typical timescale close to the one of Dy. We will model its angular momentum transfer or spin current by an STT pulse during a typical time scale of 1 ps in micromagnetic calculations starting from a demagnetized state (a prerequisite according to our experiments shown in figure 2). At the end of the spin polarized current pulse, we let the micromagnetic system evolve without any additional key ingredient (no evolution of anisotropy or saturation magnetization with time, no thermal effect, etc…). In future work, simulations with time-dependent temperature profiles which approach Tc should be carried to account for the much larger fluctuations at high temperature as well as the cooling dynamics. Nevertheless, the results summarized in figure 5 highlight a possible process that a small asymmetry to short times leads to a deterministic saturated state. In this figure, white (respectively black) domains correspond to normalized magnetization perpendicular to film plane equal to 1 (respectively -1).

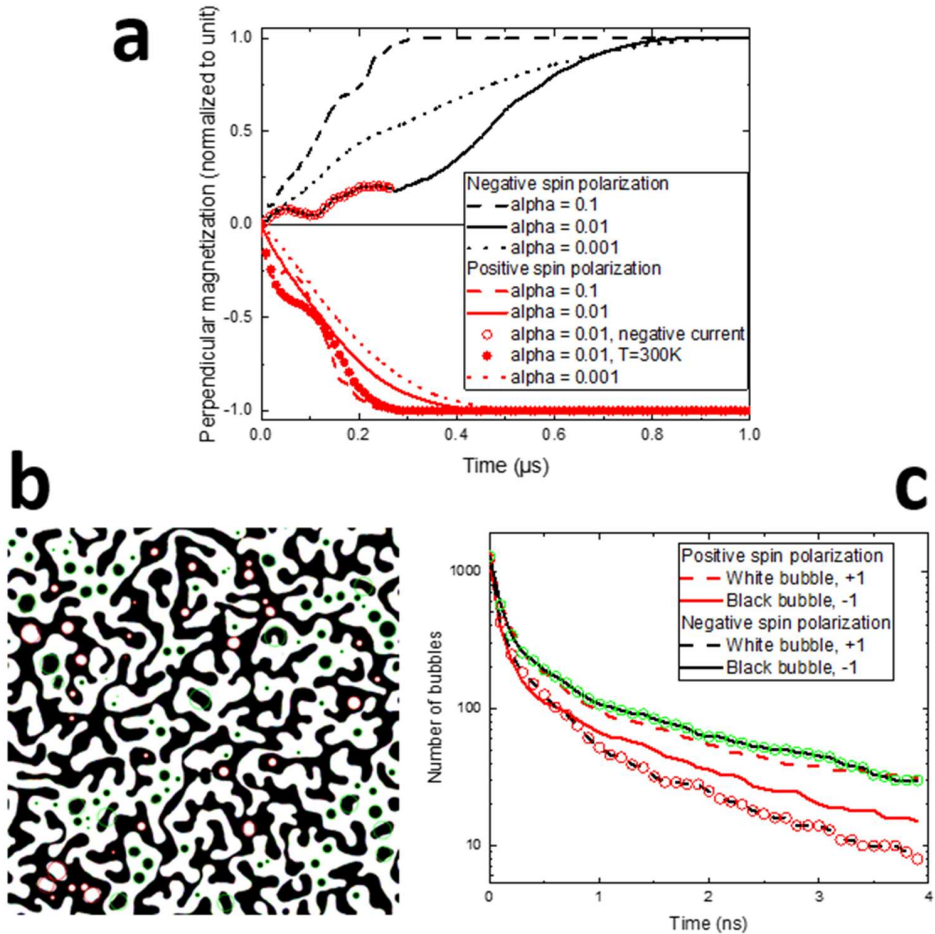

**Figure 5:** (a) Out-of-plane normalized magnetization versus time for different micromagnetic parameters : black (respectively red) curves correspond to negative (respectively positive) spin polarization; dashed lines correspond to $\alpha$=0.1, continuous lines correspond to $\alpha$=0.01 and dotted lines to $\alpha$=0.001; Red circles correspond to positive spin polarisation, negative current and $\alpha$=0.01; Red disks correspond to positive spin polarisation, positive current, $\alpha$=0.01 and T=300K. (b) Domain configuration calculated for seed 24563 and the analysis of the bubble configuration and evolution as a function of time (c).

A first batch of simulations was based on an initial random distribution of magnetization generated by Mumax3 [Mum14] with a cell size of 3 nm and results are reported in figure 5(a). It clearly shows that, independently of the material damping, the final magnetic configuration is deterministically induced by the spin polarization of the current, with magnetization opposite to the sign of the polarization of the current. We checked that by changing the sign of the current, the opposite configuration is reached (figure 5a, red circles). Finally, adding some thermal fluctuations with a calculation at 300 K does not change those conclusions (figure 5a, red disks). As can be seen, the remagnetization time is in agreement with our experimental findings. In order to check how the process is robust against initial demagnetized state, further calculations have been performed using different demagnetized state induced by other seeds of randomness. Since damping of CoHo is expected to be high, in the range of 0.1 [Wol09], we concentrated the calculations on damping values of 0.1 and 0.01. In the first 20 ns time scale, the behaviors are identical independently on the used seed (figure S4). However, the long-time calculations can be affected by periodic conditions when the sizes of the domains formed become comparable to the replicated 6 µm elementary cell. Indeed, it is clearly seen that depending on the seed used, an artificial periodic configuration stabilizes without saturation. We also check the effect of the pulse length with a fixed seed with damping of 0.1. One can clearly see that some stochasticity appears for pulse duration less than 0.4 ps meaning that the imprint of the ultra spin current / angular moment transfer has to be long enough to toggle the total magnetization (Figure S5).

In the following, we concentrate the analysis on one seed in the 30 ns time scale for which all the seeds show a similar behavior. During the remagnetization process, bubbles are formed, either white (respectively black) in a black (respectively white) background as reported in a typical configuration in Figure 5(b). The domain configurations are given in figure S6. The number of white (red circles) and black (green circles) bubbles has been counted as a function of time. The results are presented in figure 5(c). It can clearly be seen that with a positive (respectively negative) spin polarized current, an excess of white (respectively black) bubbles are formed. Those bubbles, because of their small size, are unstable and will disappear over time. As a result, with a positive (respectively negative) spin polarized current, the size of the black (respectively white) domains will increase at the expense of the white (respectively black) domains leading to a full black (respectively white) saturated state. The reversal process relies solely on the priority disappearance of the smallest domains for a gain in demagnetizing field energy. In order to show that the local demagnetization field is the driving force for the reversal, we also made a calculation removing the periodic boundary conditions to simulate the full film. We clearly see that in the first 3 ns, the distribution of the bubbles versus time is the same (figure S7).

According to our experiments, $Co_xHo_{1-x}$ alloy with concentration close and slightly above the compensation of the $Co_xHo_{1-x}$ alloy can be perfectly switched using single laser pulse over 30000 pulses with thickness of 3 nm. Therefore, the fluence has to reach full demagnetization of Co and in plane magnetization has to be avoided.

### Optical switching of $Co_xDy_{1-x}$ alloys

With those ingredients in mind, we prepared Glass/Ta(5)/Pt(5)$Co_xDy_{1-x}$(3)Ta(5)Pt(1) multilayers (thickness in nm). In case of Dy alloys, the compensation composition at room temperature was shown to be between 75% and 78% of Co

[Sen22][Fer17][Had16]. Therefore, CoxDy1-x alloys with x ranging from 80.4% to 83.3% of Co have been tested (composition slightly higher than the expected compensation at room temperature). As shown in figure S8, all samples have PMA and all optical switching could be observed. However, the most well-defined reversed domain is obtained for 81% of Co. For this concentration, we studied more deeply the reversal process which is reported in figure S9. The pulse duration / fluence state diagram and the dynamics of reversal show very similar features as those of the Co75.7Ho24.3 alloy. Here again, the reversal could not be seen in the time-resolved measurements up to 1 ns and no sign of in plane reorientation could be evidenced (See figure S9). This constitutes the first demonstration of full toggle reversal of CoxDy1-x alloys and contrasts with the one reported in [Hu23].

## Conclusions

In this study, we establish single pulse All Optical Helicity-Independent Switching in CoDy and CoHo alloys extending the number of material combinations showing this fascinating property. The reversal process, of a third type, contrasts with the first type linked to the ultrafast angular momentum transfer of (FeCo)xGd1-x ferrimagnetic alloy [Rad11,Ost12] with a reversal time of 1 ps or with the second type with precessional laser induced reversal reported in [Pen23, Sal23, Mis23] with a typical reversal time of 100 ps [Mis23]. It completes the exceptional panorama of reversal types induced by a single laser pulse. It opens up new perspectives, since it appears that, starting from a demagnetized state, any excitation leading to the appearance of a asymmetry before remagnetization can induce a deterministic reversal of magnetization on a typical time scale of less than a microsecond. It could explain the results reported by Zhang et al [Zha24] and Petty Gweha Nyoma et al [Pet24] in which dustening either Co-RE (Re=Tb, Ho, Dy) or pure Co with Gd led to single pulse AOS. Unfortunately, the reversal dynamics in case of dustening of Gd was not reported. In line with our findings, it should be slow in comparison to high Gd concentration. In all those cases including ours, the angular moment transfer appears just after the laser pulse by the demagnetization of a material subnetwork. In the case of [Yam22], the asymmetry was caused by a second pulse that brought the light angular momentum during the Co/Pt remagnetization. In the future, we forecast that many other combinations of laser induced demagnetized materials, combined with a delayed excitation, will show this third kind of magnetization reversal.

## Acknowledgements

This work is supported by the ANR-17-CE24-0007 UFO project and ANR-23-CE30-0047 SLAM project, the interdisciplinary project LUE "MAT-PULSE", part of the French PIA project "Lorraine Université d'Excellence" reference ANR-15-IDEX-04-LUE. This work was supported by the ANR through the France 2030 government grants EMCOM (ANR-22-PEEL-0009), PEPR SPIN - TOAST (ANR-22-EXSP 0003), PEPR SPIN - SPINTHEORY (ANR-22-EXSP 0009) and PEPR SPIN – SPINMAT ANR-22-EXSP- 0007. High Performance Computing resources were partially provided by the EXPLOR centre hosted by the University of Lorraine (Project: 2020M4XXX1952).

# Supplementary informations

### Micromagnetic calculations

Micromagnetic calculations have been performed using Mumax3. A cell size 3 nm × 3 nm × 9 nm has been used with a grid size of 2000×2000×1. Periodic conditions have been considered to simulate an infinite full film (PBC set to (10,10,0)). The magnetic parameters have been set up to $M_{sat}$ = 400 kA/m, $A_{ex}$ = 15×10$^{-12}$ J/m and $K_{u1}$ = 1×10$^6$ J/m$^3$ perpendicular to film plane. The damping parameter has been varied and in most simulations T = 0K. The starting configuration has been set to random using the RandomMagSeed() function. The STT torque was applied considering a fixed fictive bottom electrode with p vector along z with parameters $\lambda$=1, $\varepsilon'$ = 0 and J along z with value 20×10$^{10}$ A/m$^2$. The pulse duration was fixed to 1 ps. The polarization was fixed to 1, either positive or negative.

### Hall cross fabrication

The process begins with spin coating S1803 photoresist onto the thin film sample, followed by soft baking at 100°C for 1 minute. The first lithography step is then performed to define the Hall cross structure. RF Ar ion milling is used to etch the Hall cross. A second lithography step is carried out to shape the electrodes, followed by thermal evaporation deposition of Ti (10 nm) / Au (100 nm) for electrode formation.

# Figure S1 – Polar Kerr hysteresis loops versus Co concentration

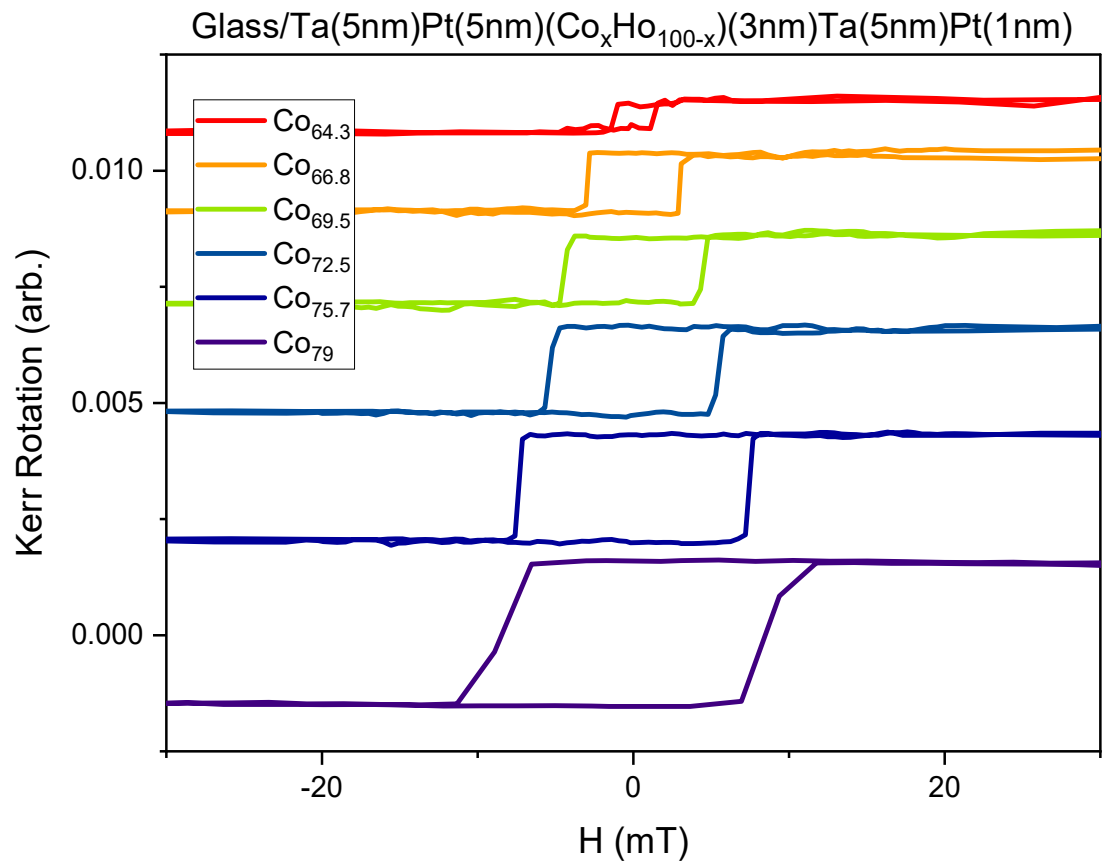

# Figure S2 – AOS versus Co concentration

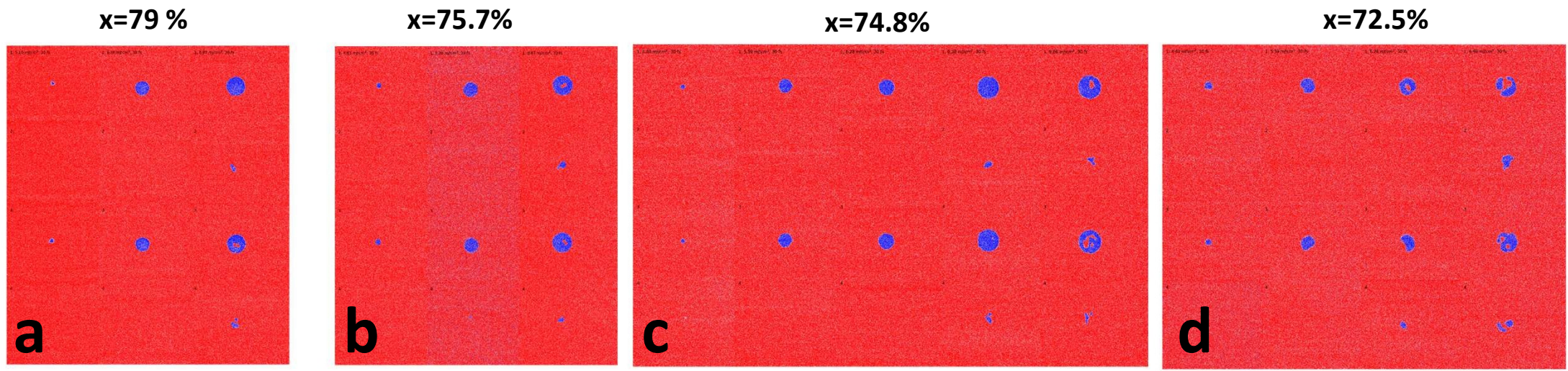

a  x=79 %
b  x=75.7%
c  x=74.8%
d  x=72.5%

# Figure S3a – AOS for 1200fs pulse duration

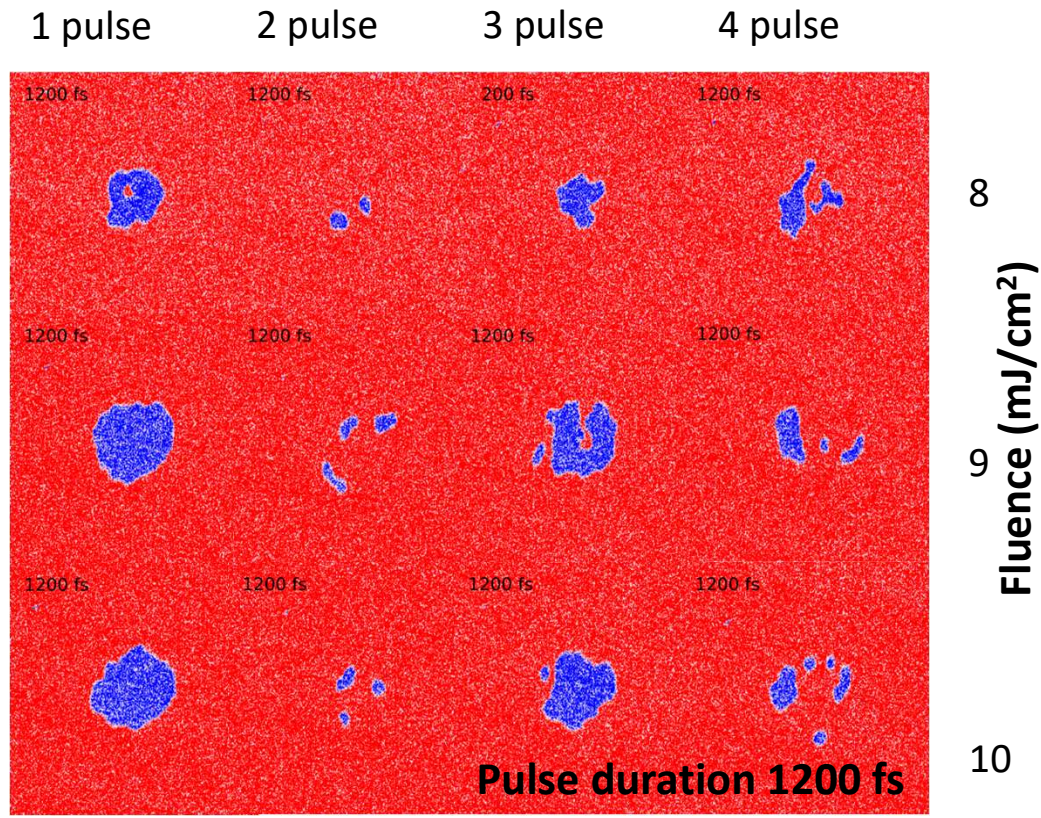

# Figure S3b – AOS for 1400fs pulse duration

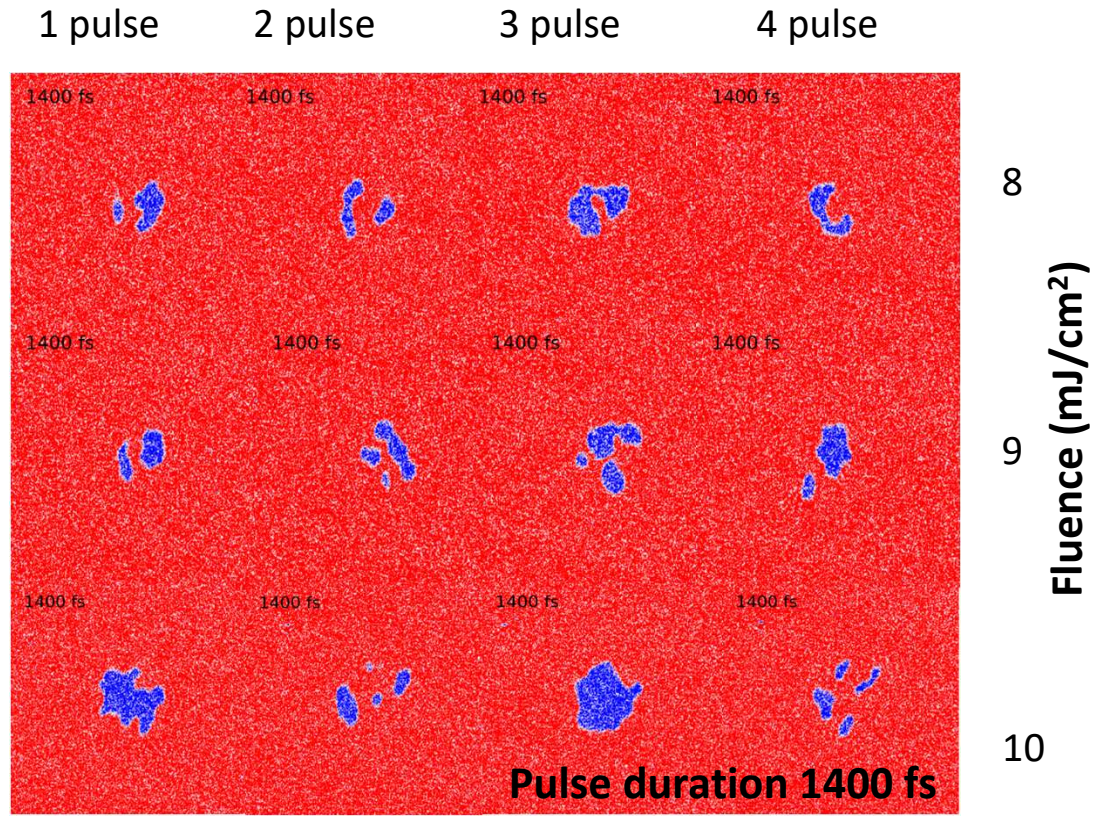

# Figure S4 – Other seeds – numerical problems

### Seed 20

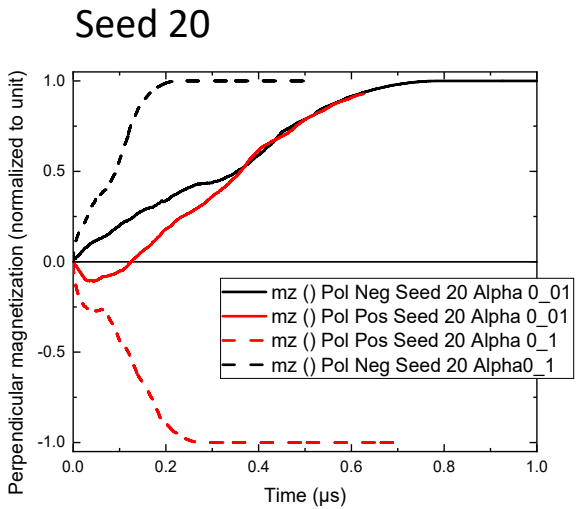

### Seed 6563

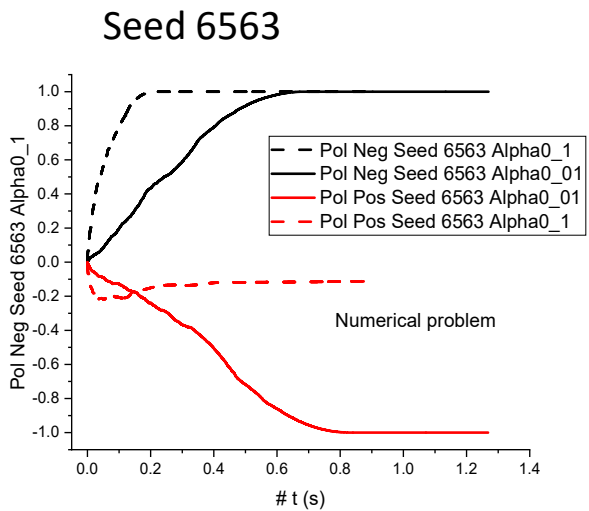

### Seed 3200

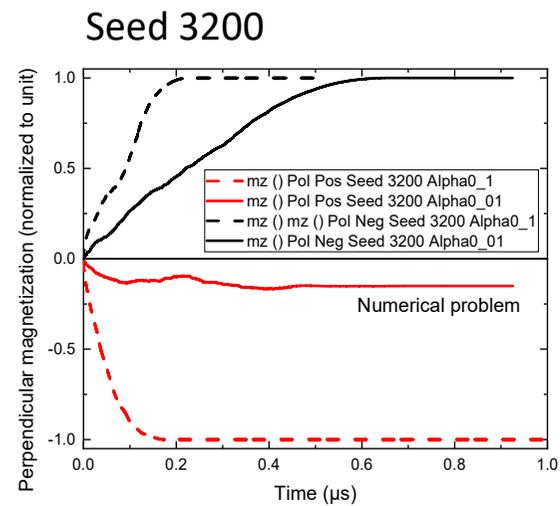

### Seed 12563

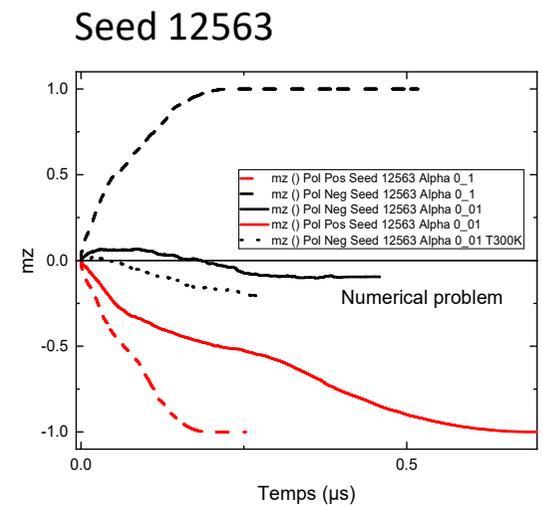

# Figure S5 – Micromagnetic process – SEED 24563

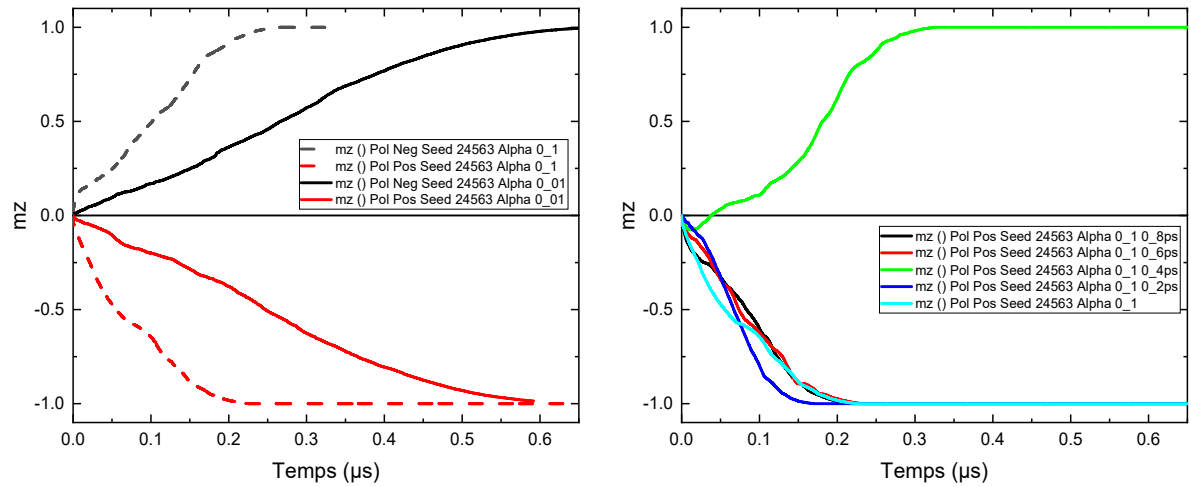

Computation underway for 3, 2, 1.5 and 0.5 ps

# Figure S6 – Micromagnetic process

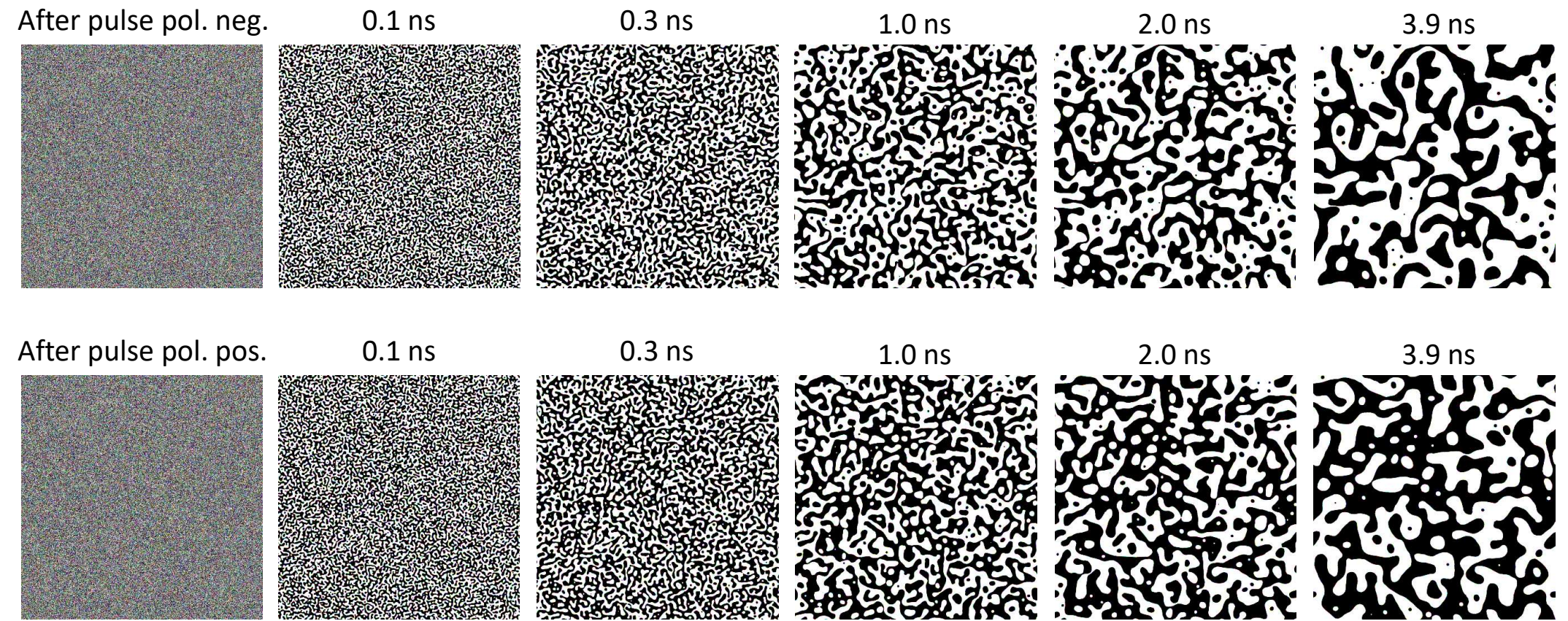

AOS_400_9nm_STT_PolNegSeed24563Alpha0_1fin.

# Figure S7 – With versus without periodic condictions

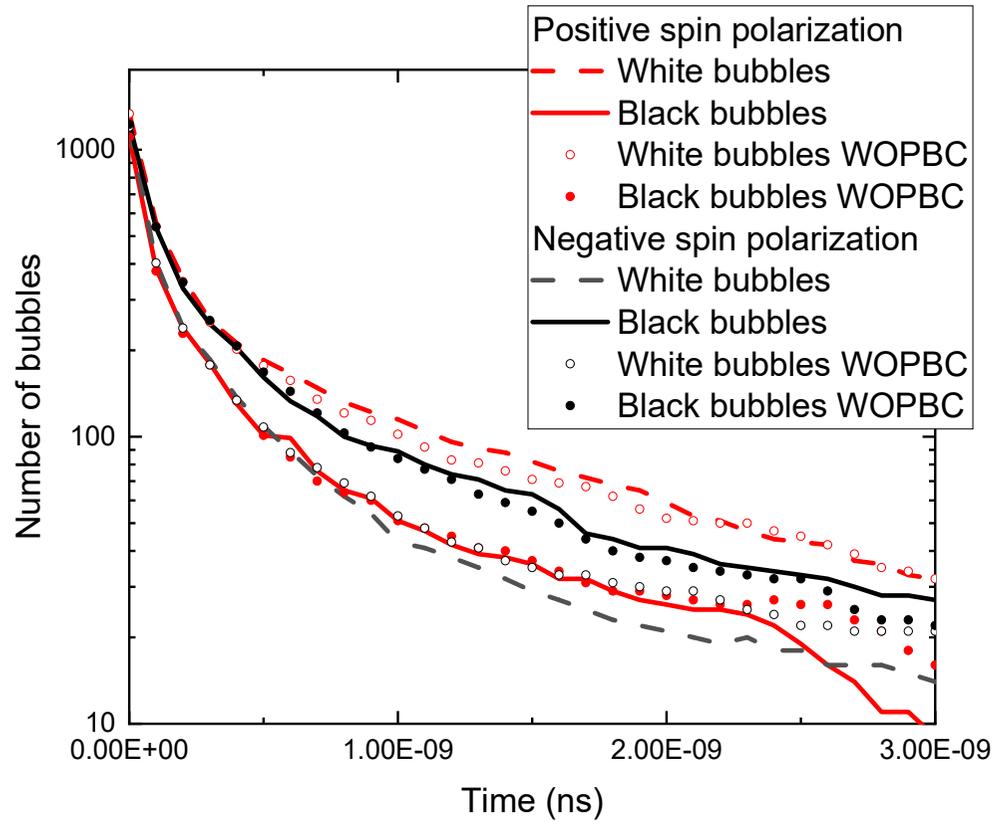

# Figure S8 – CoDy

**a**

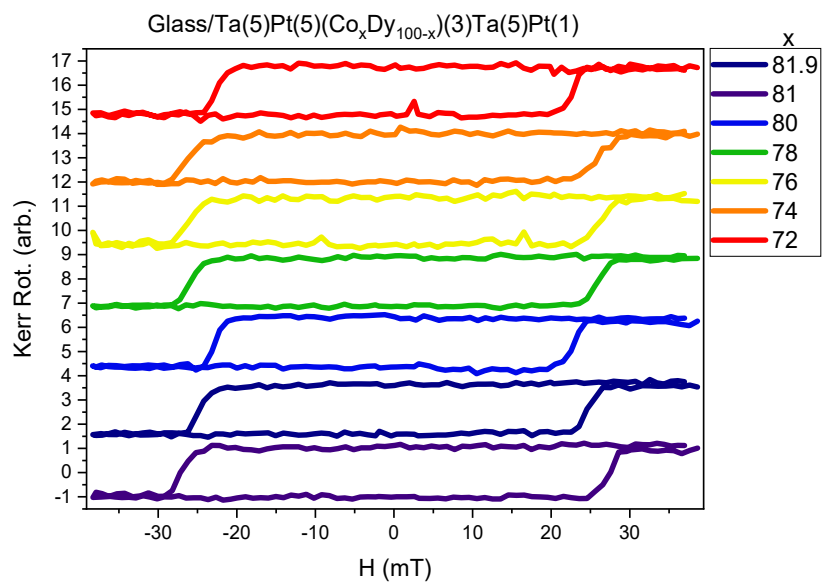

**b** Glass/Ta(5)Pt(5)(Co$_{81}$Dy$_{19}$)(3)Ta(5)Pt(1)

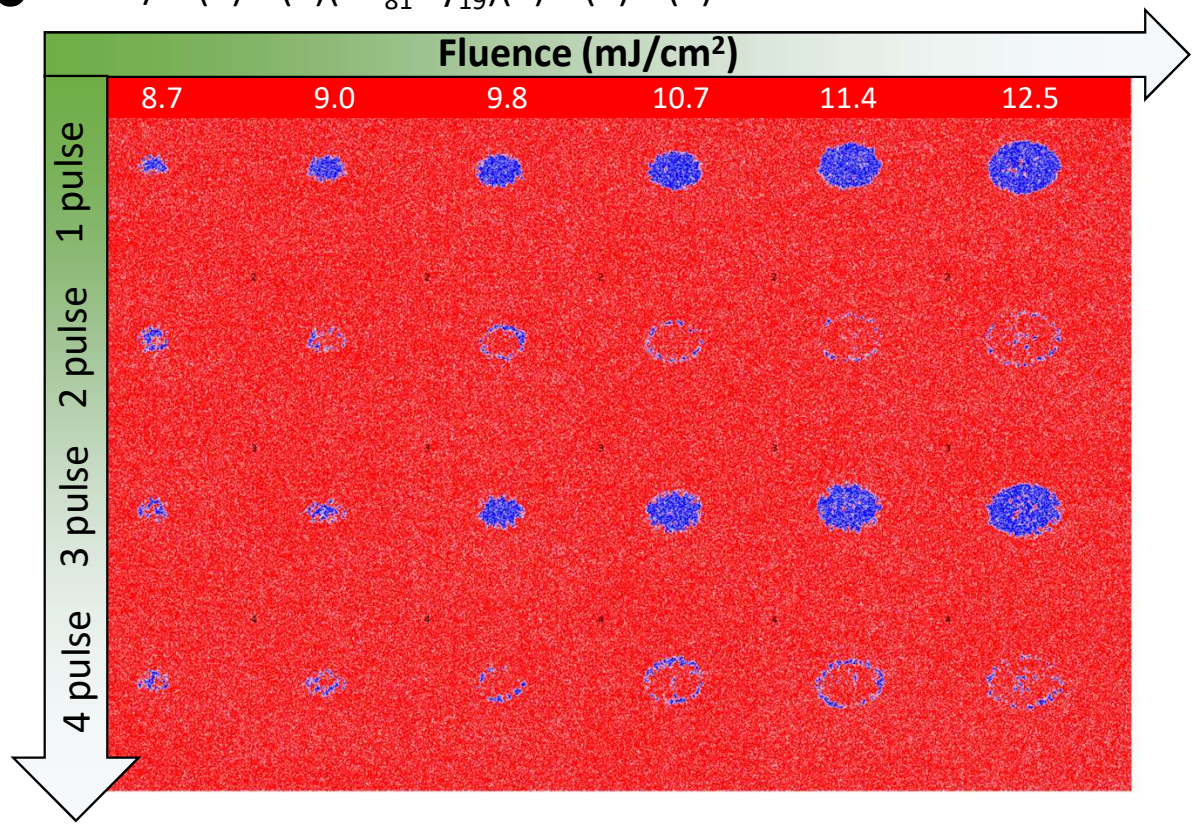

# Figure S9 – CoDy

**a** 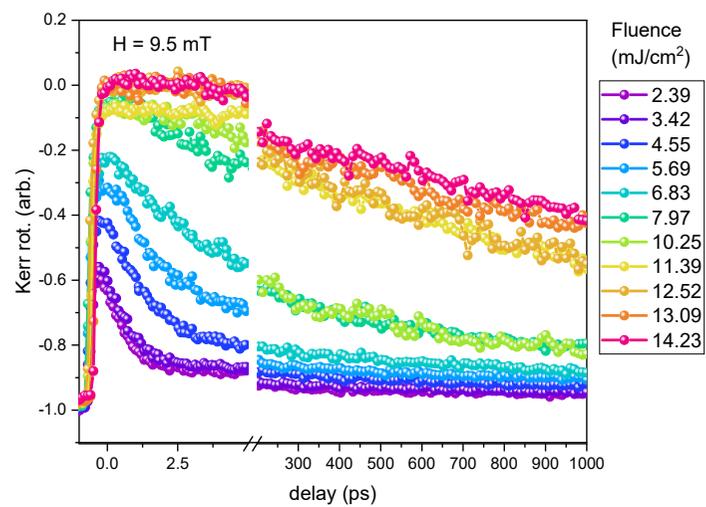

**b** 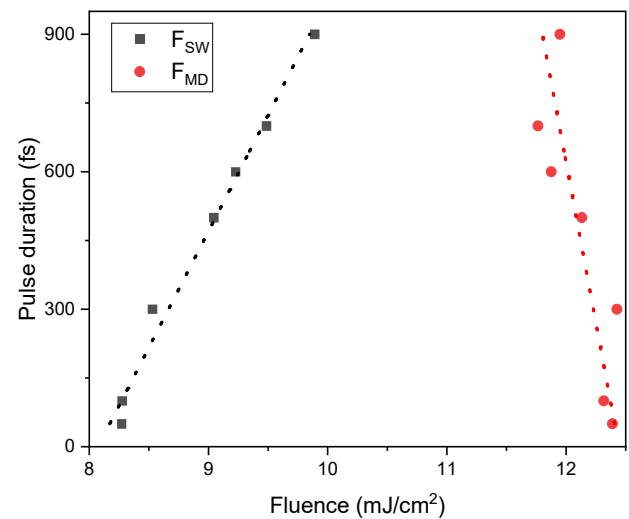

Pulse length : 50 fs